\documentclass[12pt]{article}
\usepackage{epsfig}
\title{Hadronic Atoms and Effective Interactions}
\author{Barry R. Holstein\\
Institut f\"{u}r Kernphysik\\
Forschungszentrum J\"{u}lich\\
D-52425 J\"{u}lich, Germany\\
and\\
Department of Physics and Astronomy\\
University of Massachusetts\\
Amherst, MA  01003}
\begin{document}
\begin{titlepage}
\maketitle
\begin{abstract}
We examine the problem of hadronic atom energy shifts using the
technique of effective interactions and demonstrate equivalence
with the conventional quantum mechanical approach.
\end{abstract}
\vfill
$^*$Research supported in part by the National Science Foundation
\end{titlepage} 

\section{Introduction}

In a recent paper Kong and Ravndal examined the problem of the decay
width of the pionium
atom---$\pi^+\pi^-$---using methods of effective field theory.\cite{rk1}
Specifically they employed a quasi-local approximation to the pi-pi
interaction and evaluated the width using simple second order
perturbation theory to obtain the canonical answer\cite{conv} 
\begin{equation}
\Gamma^{(0)}={8\pi\over m_\pi}k_1|a_{12}|^2|\Psi(0)|^2
\end{equation}
in terms of the
$\pi^+\pi^-\rightarrow\pi^0\pi^0$ scattering length $a_{12}$.  Here
$k_1=\sqrt{2m_{\pi^0}(m_{\pi^+}-m_{\pi^0})}\equiv\sqrt{2m_{\pi^0}\Delta
m_\pi}$ is the center of mass momentum in the final
state---$\pi^0\pi^0$---system and $|\Psi(0)|^2=m_{\pi^+}^3\alpha^3/
8\pi$ is the wavefunction at the origin.  In a subsequent
paper they examined the problem of pp scattering within a effective
local interaction picture and have shown how this matches onto 
the usual low nuclear physics description of the process.\cite{rk2}
We demonstrate below how these two discussions can be merged---by
determining the bound state energy via the presence of a pole in the
scattering matrix---and thereby make contact with the traditional
quantum mechanical discussion of hadronic atom energy shifts.\cite{qm}  In the
next section we briefly review the usual quantum mechanics
approach to the problem, while in section 3 we examine the same
problem within the effective field theory procedure and demonstrate
equivalence to the quantum mechanical results.  In section 4 we
discuss applications to pionium and to the $\pi^-p$ atom.  
Finally, in section 5 we present a short summary.

\section{Hadronic Atom Energy Shifts: Quantum Mechanical Approach}

The problem of calculating strong interaction energy shifts in
hadronic atoms is an old and well-known one, and the methods by which
to approach the subject are fairly standard.\cite{qm}  
In cases such as light pionic
atoms or pionium the system is essentially nonrelativistic so that
simple quantum mechanics can be employed.  However, before considering
such systems, we first examine the closely related problem of
scattering of particles having the {\it same} charge, as considered by
Kong and Ravndal.\cite{rk2}

Consider then a system consisting of a pair of particles $A,B$ both
having charge $+e$ and with reduced mass $m_r$.  Also, suppose 
that there exists no coupling to any other channel.  First neglect the
Coulombic interaction and consider only strong scattering.  
For simplicity, we represent the strong potential between the particles in
terms of a simple square well of depth $V_0$ and radius $R$---
\begin{equation}
V(r)=\left\{\begin{array}{cc}
-V_0 & r\leq R \\
0 &  r>R
\end{array}\right.
\end{equation}
Considering, for simplicity, S-waves the wavefunction in the interior and 
exterior regions can be written as 
\begin{equation}
\psi(r)=\left\{\begin{array}{lc}
Nj_0(Kr) & r\leq R \\
N'(j_0(kr)\cos\delta_0-n_0(kr)\sin\delta_0) & r>R
\end{array}\right.
\end{equation}
where $j_0,n_0$ are spherical harmonics and 
the interior, exterior wavenumbers are given by
$k=\sqrt{2m_rE}$, $K=\sqrt{2m_r(E+V_0)}$ respectively.  The connection
between the two forms can be made by matching logarithmic derivatives,
which yields the result
\begin{equation}
k\cot\delta_0=-{1\over R}\left[1+{1\over KRF(KR)}\right]\quad {\rm with} 
\quad F(x)=\cot x-{1\over x} 
\end{equation}
Making an effective range expansion
\begin{equation}
k\cot\delta_0=-{1\over a_0}+\ldots
\end{equation}
we find an expression for the scattering length
\begin{equation}
a_0= R\left[1-{\tan(K_0R)\over K_0R}\right]\quad 
{\rm where}\quad K_0=\sqrt{2m_rV_0}
\end{equation}
For later use we note that this can be written in the form
\begin{equation}
a_0={m_r\over 2\pi}(-{4\over 3}\pi R^3V_0)+{\cal O}(V_0^2)\label{eq:cc}
\end{equation}
The corresponding scattering amplitude is 
\begin{equation}
f(k)=e^{i\delta_0}{\sin\delta_0\over k}={1\over
k\cot\delta_0-ik}={1\over -{1\over a_0}-ik}+\ldots\label{eq:zz}
\end{equation}

Now restore the Coulomb interaction in the exterior region.  
The analysis of the
scattering proceeds as above but with the replacement of the exterior spherical
Bessel functions by appropriate Coulomb wavefunctions $F_0^+,G_0^+$
\begin{equation}
j_0(kr)\rightarrow F_0^+(r),\qquad n_0(kr)\rightarrow G_0^+(r)
\end{equation}
whose explicit form can be found in reference \cite{bj}.  For our
purposes we require only the form of these functions in the limit $kr<<1$---
\begin{eqnarray}
F_0^+(r)&\stackrel{kr<<1}{\longrightarrow}&C(\eta_+(k))(1+
{r\over a_B}+\ldots)\nonumber\\
G_0^+(r)&\stackrel{kr<<1}{\longrightarrow}&-{1\over C(\eta_+(k))}\left\{{1\over
kr}\right.\nonumber\\
&+&\left.2\eta_+(k)\left[h(\eta_+(k))+2\gamma-1+\ln {2r\over
a_B}\right]+\ldots\right\}\nonumber\\
\quad
\end{eqnarray}
Here $\gamma=0.577215..$ is the Euler constant,
\begin{equation}
C^2(x)={2\pi x\over \exp(2\pi x)-1}
\end{equation}
is the usual Coulombic enhancement factor, $a_B=1/m_r\alpha$ is the Bohr
radius, $\eta_+(k)=1/ka_B$, and 
\begin{equation}
h(\eta_+(k))={\rm Re}H(i\eta_+(k))=
\eta_+^2(k)
\sum_{n=1}^\infty {1\over n(n^2+\eta_+^2(k))}-\ln\eta_+(k)-\gamma
\end{equation}
where $H(x)$ is the analytic function
\begin{equation}
H(x)=\psi(x)+{1\over 2x}-\ln(x)
\end{equation}
Equating interior and exterior logarithmic derivatives we find
\begin{eqnarray}
KF(KR)&=&{\cos\delta_0 {F_0^+}'(R)-\sin\delta_0{G_0^+}'(R)\over
\cos\delta_0F_0^+(R)
-\sin\delta_0G_0^+(R)}\nonumber\\
&=&{k\cot\delta_0C^2(\eta_+(k)){1\over a_B}-{1\over R^2}\over
k\cot\delta_0C^2(\eta_+(k))+{1\over R}+{2\over
a_B}\left[h(\eta_+(k))-\ln{a_B\over 2R}+2\gamma-1\right]}\nonumber\\
\quad\label{eq:aa}
\end{eqnarray}
Since $R<<a_B$ Eq. \ref{eq:aa} can be written in the form
\begin{equation}
k\cot\delta_0C^2(\eta_+(k))+{2\over
a_B}\left[h(\eta_+(k))-\ln{a_B\over 2R}+2\gamma-1\right]\simeq -{1\over a_0}
\end{equation}
The scattering length $a_C$ in the presence of the Coulomb interaction
is conventionally defined as\cite{pres}
\begin{equation}
k\cot\delta_0C^2(\eta_+(k))+{2\over a_B}h(\eta_+(k))=-{1\over a_C}
+\ldots\label{eq:dd}
\end{equation}
so that we have the relation
\begin{equation}
-{1\over a_0}=-{1\over a_C}-{2\over a_B}(\ln{a_B\over 2R}+1-2\gamma)\label{eq:yy}
\end{equation}
between the experimental scattering length---$a_C$---and that which would
exist in the absence of the Coulomb interaction---$a_0$.  

As an aside we note that $a_0$ is not itself an
observable since the Coulomb interaction {\it cannot} be turned off.
However, in the case of the pp interaction isospin invariance
requires $a_0^{pp}=a_0^{nn}$ so that one has the prediction
\begin{equation}
-{1\over a_0^{nn}}=-{1\over a_C^{pp}}-\alpha M_N(\ln{1\over \alpha M_NR}+1
-2\gamma)\label{eq:oo}
\end {equation}
While, strictly speaking, this is a model dependent result, Jackson
and Blatt have shown by treating the interior
Coulomb interaction perturbatively that a version of this result 
with $1\rightarrow 0.824$ is approximately valid
for a wide range of strong interaction potentials\cite{bj} and the
correction indicated in Eq. \ref{eq:oo} 
is essential in restoring agreement between the widely
discrepant---$a_0^{nn}=-18.8$ fm vs. $a_C^{pp}=-7.82$ fm---values
obtained experimentally.  

Returning to the problem at hand, the 
experimental scattering amplitude can then be written as
\begin{eqnarray}
f_C^+(k)&=&{e^{2i\sigma_0}C^2(\eta_+(k))\over -{1\over a_C}-{2\over
a_B}h(\eta_+(k))-ikC^2(\eta_+(k))}\nonumber\\
&=&{e^{2i\sigma_0}C^2(\eta_+(k))\over
-{1\over a_C}-{2\over a_B}H(i\eta_+(k))}\label{eq:hh}
\end{eqnarray}
where $\sigma_0={\rm arg}\Gamma(1-i\eta_+(k))$ is the Coulomb phase. 

Analysis of the situation involving particles of {\it opposite} charge
is similar, except that in this case in the absence of strong
interaction effects there will exist, of course, Coulomb bound states
at momentum $k_n=i\kappa_n=i/na_B$ and energy $E_n=-\kappa_n^2/2m_r
=-m_r\alpha^2/2n^2$ with $n=1,2,3,\ldots$  In the presence of strong
interactions between these particles, however, the energies will be
shifted.  One approach to the calculation of this shift is to examine
the corresponding scattering process $A+B\rightarrow A+B$.  Then 
the existence of a bound state is indicated by the presence of a
pole along the positive imaginary axis---{\it i.e.} for $\kappa>0$ 
under the analytic continuation $k\rightarrow i\kappa$.
(If we (temporarily) neglect Coulomb effects and look only at
strong scattering, this occurs when $\kappa=1/a_0$ in the
case of Eq. \ref{eq:zz}.)

However, in the case of oppositely charged particles $A,B$ the
analysis of the scattering amplitude must be in terms of appropriate
Coulomb wavefunctions.  If, as before, we include Coulomb
effects only in the exterior region, then the appropriate forms of the
wavefunctions for $kr<<1$ are given in ref.\cite{true} as
\begin{eqnarray}
F_0^-(r)&\stackrel{kr<<1}{\longrightarrow}& C(\eta_+(k))\left(1-{r\over
a_B}+\ldots\right)\nonumber\\
G_0^-(r)&\stackrel{kr<<1}{\longrightarrow}& -{1\over
C(\eta_+(k))}\left\{{1\over kr}\right.\nonumber\\
&-&\left.2\eta_+(k)\left[H(i\eta_+(k))-i{\pi\over
2}\coth(\pi\eta_+(k))+2\gamma-1+\ln{i{2r\over
a_B}}\right]+\ldots\right\}\nonumber\\
\quad
\end{eqnarray}
Equating interior and exterior logarithmic deriviatives as before, we
find
\begin{eqnarray}
KF(KR)&=&{\cos\delta_0{F_0^-}'(R)-\sin\delta_0{G_0^-}'(R)
\over \cos\delta_0F_0^-(R)-\sin\delta_0G_0^-(R)}\nonumber\\
&=&{-k\cot\delta_0C^2(-\eta_+(k)){1\over a_B}-{1\over R^2}\over 
k\cot\delta_0C^2(-\eta_+(k))+{1\over R}-{2\over
a_B}(h(\eta_+(k))-\ln{a_B\over 2R}+2\gamma-1)}\nonumber\\
\quad
\end{eqnarray}
Thus we have 
\begin{equation}
k\cot\delta_0C^2(-\eta_+(k))-{2\over
a_B}\left[h(\eta_+(k))-\ln{a_B\over 2R}+2\gamma-1\right]\simeq
-{1\over a_0}
\end{equation}
In this case the scattering length $a_C$ in the presence of the Coulomb
interaction is defined as
\begin{equation}
k\cot\delta_0C^2(-\eta_+(k))-{2\over a_B}h(\eta_+(k))=-{1\over a_C}
\end{equation}
so that we have the relation
\begin{equation}
-{1\over a_0}=-{1\over a_C}+{2\over a_B}(\ln{a_B\over 2R}+1-2\gamma)
\end{equation}
The corresponding scattering amplitude is then
\begin{eqnarray}
f_C^-(k)&=&{e^{-2i\sigma_0}C^2(-\eta_+(k))\over -{1\over a_C}+
{2\over a_B}h(\eta_+(k))-ikC^2(-\eta_+(k))}\nonumber\\
&=&{e^{-2i\sigma_0}C^2(-\eta_+(k))\over -{1\over a_C}+{2\over a_B}
[H(i\eta_+(k))-i\pi\coth\pi\eta_+(k))]}\label{eq:hhh}
\end{eqnarray}
Under the continuation $k\rightarrow i\kappa$ we have
\begin{equation}
H(i\eta_+(k))-i\pi\coth(\pi\eta_+(k))\rightarrow H(\xi)
+\pi\cot\pi\xi\label{eq:ww}
\end{equation}
where $\xi=1/\kappa a_B$.  The existence of a bound state is then
signalled by 
\begin{equation} 
-{1\over a_C}+{2\over a_B}(H(\xi)+\pi\cot\pi\xi)=0\label{eq:bb}
\end{equation}
In the limit of no strong interaction---$a_C\rightarrow 0$---we
find $\xi_n=1/\kappa_na_B=n$ and the usual Coulomb bound state
energies
\begin{equation}
E_n=-{\kappa_n^2\over 2m_r}=-{m_r\alpha^2\over 2n^2},\quad n=1,2,3,\ldots
\end{equation}
while if $a_C\neq 0$ there exists a solution to Eq. \ref{eq:bb}
\begin{equation} 
\xi={1\over \kappa_na_B}\approx
n+{2a_C\over a_B}
\end{equation} 
and a corresponding energy shift
\begin{equation}
\Delta E_n=-E_n{4a_C\over na_B}+{\cal O}({a_C\over a_B})^2\label{eq:ff}
\end{equation}
which is the conventional result.\cite{true},\cite{des}  

It is important to note here that Eq. \ref{eq:ff} is written in a form that
relates one {\it experimental} quantity---the energy shift $\Delta E_n$---to
another---the scattering length $a_C$.  Hence it is {\it
model-independent}, even though, for clarity, we have employed a
particular model in its derivation.  This feature means that it is
an ideal case for an effective interaction approach, as will be shown
in the next section.  However, we first complete our quantum
mechanical discussion.

An alternative approach involves the use of bound state perturbation 
theory.\cite{rk1}  In this case
the problem simplifies because of the feature that the range of the
strong interaction---$R$---is much less than the Bohr radius---$a_B$.
Thus we may write\cite{des}
\begin{eqnarray}
\Delta E_n&=&<\Psi|V|\Psi>+\sum_{n\neq
0}{<\Psi|V|n><n|V|\Psi>\over E_n-E_0}+\ldots\nonumber\\
&\simeq& |\Psi(0)|^2\times \lim_{k\rightarrow i\kappa}<\phi_f|V|\psi_i^{(+)}> 
\end{eqnarray}
Connection with the scattering length may be made via\cite{wf}
\begin{equation}
f(k)=-{m_r\over 2\pi}<\phi_f|V|\psi_i^{(+)}> \simeq -a
\end{equation}
so that for weak potentials we have
\begin{equation}
a\approx -{m_r\over 2\pi}\int d^3rV(r)=-{m_r\over 2\pi}{4\over 3}\pi R^3V_0
\end{equation}
in agreement with Eq. \ref{eq:cc}.
Then since for Coulombic wavefunctions
\begin{equation}
|\Psi(0)|^2={1\over \pi n^3 a_B^3}
\end{equation}
we have
\begin{equation}
\Delta E_n= {2\pi a\over m_r}|\Psi(0)|^2+\ldots=-E_n{4a\over na_B}+\ldots
\end{equation}
as found via continuation of the scattering amplitude.  (However, in
this form it is not completely clear whether the relevant scattering length is
the experimental quantity $a_C$ or its Coulomb subtracted analog $a_0$.)

In the real world, of course, this simple model is no longer valid, since
realistic hadronic atoms, such as pionium or $\pi^-p$, are coupled to
unbound systems---$\pi^0\pi^0$ or $\pi^0n,\gamma n$---and must be treated as 
multi-channel problems.  Nevertheless the methods generalize
straightforwardly from those given above.  Specifically, in the
absence of Coulomb interactions the
scattering amplitude for the $\pi\pi$ system 
can be given in the two-channel K-matrix form\footnote{There exists
also a coupling of $\pi^+\pi^-$ via the anomaly to the $\pi^0\gamma$ channel,
but this is p-wave and can be neglected for s-states
such as considered here.}
\begin{equation} 
f^{-1}=-A^{-1}-i\left(\begin{array}{ll}
k_1 & 0\\
0 & k_2\end{array}\right)
\end{equation}
where $k_1,k_2$ are the center of mass momenta in the 
1(neutral),2(charged) channels respectively and 
\begin{equation}
A=\left(\begin{array}{ll}
a_{11} & a_{12}\\
a_{21} & a_{22}\end{array}\right)\label{eq:rr}
\end{equation}
is a real matrix given in 
terms of the coupled channel scattering lengths.  The
inverse of the matrix $A$ is easily determined and the 
resultant form of the scattering amplitude is
\begin{eqnarray}
f&=&{1\over 1+ik_1a_{11}+ik_2a_{22}-k_1k_2{\rm det}A}\nonumber\\
&\times&\left(\begin{array}{cc}
-a_{11}(1+ik_2a_{22})+ik_2a_{12}a_{21} & -a_{12}\\
-a_{21} &
-a_{22}(1+ik_1a_{11})+ik_1a_{12}a_{21}\end{array}\right)\nonumber\\
\quad\label{eq:ss}
\end{eqnarray}
The existence of a bound state is, as before, indicated by the
presence of a pole---{\it i.e.} by the condition
\begin{equation}
1+ik_1a_{11}+ik_2a_{22}-k_1k_2{\rm det} A=0\label{eq:ee}
\end{equation}
after approporiate analytic continuation.  In the case of hadronic
atoms, we must, of course, correctly include the Coulomb effects.
From our single channel experience above, this is done via the prescription
\begin{equation}
a_{ij}\rightarrow a_{ij}^C,\quad -ik_2\rightarrow {2\over a_B}(H(\xi)+\pi\cot\pi\xi)
\end{equation}
whereby the bound state condition---Eq.
\ref{eq:ee}---reads
\begin{equation}
0=1+{2\over a_B}(H(\xi)+\pi\cot\pi\xi)\left(-a_{22}^C+{ik_1a_{12}^Ca_{21}^C
\over 1+ik_1a_{11}^C}\right)\label{eq:tt}
\end{equation} 
Thus we find 
\begin{equation}
\Delta E_n= -E_n{4\over na_B}(a_{22}^C-{ik_1a_{12}^Ca_{21}^C\over
1+ik_1a_{11}^C})
+{\cal O}({a_{ij}^C\over a_B})^2
\end{equation}
as the coupled channel generalization of Eq. \ref{eq:ff}.  The real
component of the energy shift is, of course, to lowest order identical
to the single channel result.  What is new is that there has developed
an imaginary component, corresponding to a decay width
\begin{equation}
\Gamma_n=-2{\rm Im}\Delta E_n=-E_n{8\over na_B}
k_1|a_{12}^C|^2+\ldots={4\pi\over m_r^{(2)}}k_1|a_{12}^C|^2|\Psi(0)|^2+\ldots
\end{equation}
which is precisely what is expected from Fermi's golden rule
\begin{eqnarray}
\Gamma_n&=&\int{d^3k_1\over (2\pi)^3}2\pi\delta(\Delta E-{k_1^2
\over 2m_r^{(1)}})|<\phi_f|V|\Psi_i>|^2\nonumber\\
&=&({2\pi\over \sqrt{m_r^{(1)}m_r^{(2)}}}a_{12}^C)^2{m_r^{(1)}\over
\pi}k_1|\Psi(0)|^2
\end{eqnarray}
or from second order perturbation theory.

\section{Hadronic Atom Energy Shifts:  Effective Field Theory}

Identical results may be obtained from effective field theory and in
many ways the derivation is clearer and more intuitive.\cite{kap}  First 
consider the situation that we have two particles A,B interacting only
via a local strong interaction, so that the effective Lagrangian can
be written as
\begin{equation}
{\cal L}=\sum_{i=A}^B\Psi_i^\dagger(i{\partial\over \partial
t}+{\nabla^2\over
2m_i})\Psi_i-C_0\Psi_A^\dagger\Psi_A\Psi_B^\dagger\Psi_B
+\ldots\label{eq:gg}
\end{equation}
The T-matrix is then given in terms of the multiple
scattering series shown in Figure 1
\begin{equation}
T_{fi}(k)=-{2\pi\over m_r}f(k)=C_0+C_0^2G_0(k)+C_0^3G_0^2(k)+\ldots
={C_0\over 1-C_0G_0(k)}\label{eq:ll}
\end{equation}
where $G_0(k)$ is the amplitude for particles $A,B$ to travel from zero
separation to zero separation---{\it i.e} the propagator
$D_F(k;\vec{r}'=0,\vec{r}=0)$---
\begin{equation}
G_0(k)=\lim_{\vec{r}',\vec{r}\rightarrow 0}\int{d^3s\over (2\pi)^3}{
e^{i\vec{s}\cdot\vec{r}'}e^{-i\vec{s}\cdot\vec{r}}\over {k^2\over
2m_r}-{s^2\over 2m_r}+i\epsilon}=\int{d^3s\over (2\pi)^3}
{2m_r\over k^2-s^2+i\epsilon}
\end{equation}
Equivalently $T_{fi}(k)$
satisfies a Lippman-Schwinger equation
\begin{equation}
T_{fi}(k)=C_0+C_0G_0(k)T_{fi}(k).
\end{equation}
whose solution is given in Eq. \ref{eq:ll}.

\begin{figure}[htb]
\begin{center}
\epsfig{figure=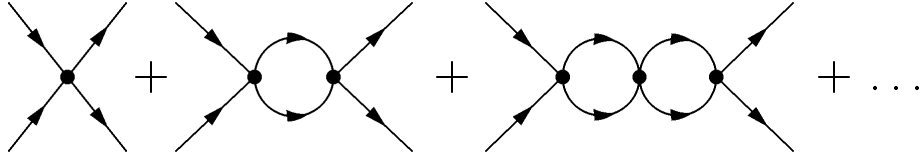,height=20mm}
\end{center}
\vspace{-8mm}
\caption{The multiple scattering series.}
\label{fig}
\end{figure}

The function $G_0(k)$ is divergent and must be defined via some sort of
regularization.  There are a number of ways by which to do this.  We
shall herein use a cutoff
regularization with $k_{max}=\mu$ we have
\begin{equation}
G_0(k)=-{m_r\over 2\pi}({2\mu\over \pi}+ik)
\end{equation}
Equivalently, one could subtract at an unphysical momentum point, as
proposed by Gegelia\cite{geg}
\begin{equation}
G_0(k)=\int{d^3s\over (2\pi)^3}({2m_r\over
k^2-s^2+i\epsilon}+{2m_r\over \mu^2+s^2})=-{m_r\over 2\pi}(\mu+ik)
\end{equation}
which has been shown by Mehen and Stewart\cite{ms} to be equivalent to
the PDS sceme of Kaplan, Savage and Wise.\cite{kap} 
In any case, the would-be linear divergence is, of course, cancelled by
introduction of a counterterm, which renormalizes $C_0$ to $C_0(\mu)$.
The scattering amplitude is then
\begin{equation}
f(k)=-{m_r\over 2\pi}\left({1\over{1\over C_0(\mu)}-G_0(k)}\right)={1\over 
-{2\pi\over m_rC_0(\mu)}-{2\mu\over \pi}-ik}
\end{equation}
Comparing with Eq. \ref{eq:zz} we identify the scattering length as
\begin{equation}
-{1\over a_0}=-{2\pi\over m_rC_0(\mu)}-{2\mu\over \pi}
\end{equation}

More interesting is the case where we restore the Coulomb interaction
between the particles.  The derivatives in Eq. \ref{eq:gg} then become
covariant and the bubble sum is evaluated with static photon exchanges
between each of the lines---each bubble is replaced by one involving a
sum of zero, one, two, etc. Coulomb interactions, as shown in Figure
2.  

\begin{figure}[htb]
\begin{center}
\epsfig{figure=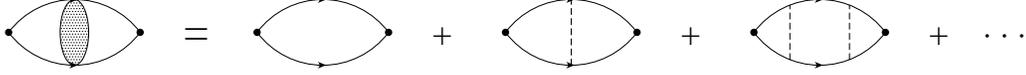,height=10mm}
\end{center}
\vspace{-5mm}
\caption{The Coulomb corrected bubble.}
\label{fig2}
\end{figure}

The net result in the case of same charge scattering is the
replacement of the free propagator by its Coulomb analog
\begin{eqnarray}
G_0(k)\rightarrow G_C^+(k)&=&\lim_{\vec{r}',\vec{r}\rightarrow 0}
\int{d^3s\over
(2\pi)^3}{\psi^+_{\vec{s}}(\vec{r}'){\psi^+_{\vec{s}}}^*
(\vec{r})\over {k^2\over 2m_r}-{s^2\over 2m_r}+i\epsilon}\nonumber\\
&=&\int{d^3s\over (2\pi)^3}{2m_rC^2(\eta_+(s))
\over k^2-s^2+i\epsilon}\label{eq:uuu}
\end{eqnarray}
where
\begin{equation}
\psi^+_{\vec{s}}(\vec{r})=C(\eta_+(s))e^{i\sigma_0}
e^{i\vec{s}\cdot\vec{r}}
{}_1F_1({-i\eta_+(s),1,isr-i\vec{s}\cdot\vec{r}})
\end{equation}
is the outgoing Coulomb wavefunction for repulsive Coulomb scattering.\cite{lal}
Also in the initial and final states the influence of static photon
exchanges must be included to all orders, which produces the factor 
$C^2(2\pi\eta_+(k))\exp(2i\sigma_0)$.  Thus the repulsive 
Coulomb scattering amplitude becomes
\begin{equation}
f_C^+(k)=-{m_r\over 2\pi}{C_0C^2(\eta_+(k))\exp 2i\sigma_0
\over 1-C_0{}G_C^+(k)}
\end{equation}
The momentum integration in Eq. \ref{eq:uuu} can be performed as before
using cutoff regularization, yielding 
\begin{equation}
G_C^+(k)=-{m_r\over 2\pi}\left\{{2\mu\over \pi}+{2\over
a_B}\left[H(i\eta_+(k))-\ln{\mu a_B\over 2\pi}-\zeta\right]\right\}
\end{equation}
where $\zeta=\ln 2\pi -\gamma$.  (Equivalently, in the unphysical momentum subtraction
scheme
\begin{eqnarray}
G_C^+(k)&=&-{m_r\over 2\pi}{2\over a_B}(H(i\eta_+(k))-H({1\over \mu
a_B}))\nonumber\\
&\simeq&-{m_r\over 2\pi}(\mu+{2\over a_B}[H(i\eta_+(k))-\ln{\mu a_B}-\psi(1)])
\end{eqnarray}
We have then
\begin{eqnarray}
f_C^+(k)={C^2(\eta_+(k))e^{2i\sigma_0}\over -{2\pi\over
m_rC_0(\mu)}-{2\mu\over \pi}-{2\over a_B}\left[H(i\eta_+(k))-\ln{\mu
a_B\over 2\pi}-\zeta\right]}\nonumber\\
={C^2(\eta_+(k))e^{2i\sigma_0}\over -{1\over a_0}-{2\over
 a_B}\left[h(\eta_+(k)-\ln{\mu a_B\over 2\pi}-\zeta\right]
-ikC^2(\eta_+(k))}
\end{eqnarray}
Comparing with Eq. \ref{eq:hh} we identify the Coulomb scattering
length as
\begin{equation}
-{1\over a_C}=-{1\over a_0}+{2\over a_B}(\ln{\mu a_B\over 2\pi}+\zeta)
\end{equation}
which matches nicely with Eq. \ref{eq:yy} if a reasonable cutoff
$\mu\sim m_\pi\sim 1/R$ is employed.  The scattering amplitude then
has the simple form
\begin{equation}  
f_C^+(k)={C^2(\eta_+(k))e^{2i\sigma_0}\over -{1\over a_C}-
{2\over a_B}H(i\eta_+(k))}
\end{equation}
in agreement with Eq. \ref{eq:hh}.

Now consider oppositely charged particles.  In this case
the analysis is parallel to that above, 
but there exist important new wrinkles in
that the intermediate state sum in the Coulomb propagator must now
include bound states
\begin{eqnarray}
G_0(k)\rightarrow G_C^-(k)&=&\lim_{\vec{r}',\vec{r}\rightarrow
0}\left[\sum_{n\ell m}{\psi_{n\ell m}(\vec{r}')\psi_{n\ell
m}^*(\vec{r})\over {k^2\over 2m_r}+{m_r\alpha^2\over
2n^2}}+\int{d^3s\over (2\pi)^3}{\psi^-_{\vec{s}}(\vec{r}')
{\psi^-_{\vec{s}}}^*(\vec{r})\over {k^2\over 2m_r}-{s^2\over
2m_r}+i\epsilon}\right]\nonumber\\
&=&{2m_r\over \pi a_B}\sum_{n=1}^\infty{\eta_+^2(k)\over  n(
n^2+\eta_+^2(k))}+\int{d^3s\over (2\pi)^3}
{2m_r C^2(-\eta(s))\over k^2-s^2+i\epsilon}\label{eq:jjj}
\end{eqnarray}
where 
\begin{equation}
\psi^-_{\vec{s}}(\vec{r})=C(-\eta_+(s))e^{-i\sigma_0}e^{i\vec{s}\cdot\vec{r}}
{}_1F_1(i\eta_+(s),1,isr-i\vec{s}\cdot\vec{r})
\end{equation}
is the outgoing Coulomb wavefunction for attractive Coulomb scattering and
\begin{equation}
\psi_{n\ell m}(\vec{r})=\left[(2a_B)^3{(n-\ell-1)!\over
2n[(n+\ell)!]^3}\right]^{1\over 2}e^{-a_Br}(2a_Br)^\ell
L_{n-\ell-1}^{2\ell+1}(2a_Br)Y^m_\ell(\theta,\phi)
\end{equation}
is the bound state wavefunction corresponding to quantum numbers
$n\ell m$ with $L^i_j(x)$ being the associated Laguerre polynomial.  
Using the identity
\begin{equation}
C^2(-\eta_+(k))=-C^2(\eta_+(k))+2\pi\eta_+(k)\coth\pi\eta_+(k)
\end{equation}
we can write Eq. \ref{eq:jjj} as
\begin{eqnarray}
G_C^-(k)&=&-G_C^+(k)+2m_r\int{d^3s\over
(2\pi)^3}{2\pi\eta_+(s)\coth\pi\eta_+(s)\over
k^2-s^2+i\epsilon}\nonumber\\
&+&{2m_r\over \pi a_B}\sum_{n=1}^\infty{\eta_+^2(k)\over
n(n^2+\eta_+^2(k))}
\nonumber\\
&=&-{m_r\over 2\pi}\left\{{2\mu\over \pi}-{2\over a_B}\left[
H(i\eta_+(k))-i\pi\coth\pi\eta_+(k)-\ln{\mu a_B\over 2\pi}-\zeta
\right]\right\}\nonumber\\
\quad
\end{eqnarray}
where the integration is done via contour methods and the 
the contribution from the hyperbolic cotangent poles precisely cancels
the bound state term.  The resulting attractive Coulomb scattering 
amplitude is given by
\begin{eqnarray}
f_C^-(k)&=&{C^2(-\eta_+(k))e^{-2i\sigma_0}\over -{2\pi\over
m_rC_0(\mu)}-{2\mu\over \pi}+{2\over a_B}\left[H(i\eta_+(k))
-i\pi\coth\pi\eta_+(k)-\ln{\mu a_B\over 2\pi}-\zeta\right]}\nonumber\\
&=&{C^2(-\eta_+(k))e^{-2i\sigma_0}\over -{1\over a_0}+{2\over a_B}
\left[h(\eta_+(k))-\ln{\mu a_B\over 2\pi}-\zeta\right]-ikC^2(-\eta_+(k))}
\end{eqnarray}
Identifying the Coulomb scattering length via
\begin{equation}
-{1\over a_C}=-{1\over a_0}-{2\over a_B}(\ln{\mu a_B\over 2\pi}+\zeta)
\end{equation}
this reduces to the simple form
\begin{equation}
f_C^-(k)={C^2(-\eta_+(k))e^{-2i\sigma_0}\over -{1\over a_C}+
{2\over a_B}[H(i\eta_+(k))-i\pi\coth\pi\eta_+(k)]}
\end{equation}
in agreement with Eq. \ref{eq:hhh}.  In order to go to the bound state limit
we can utilize the continuation
\begin{equation}
H(i\eta_+(k))
-i\pi\coth\pi\eta_+(k)\stackrel{k\rightarrow i\kappa}
{\longrightarrow} H(\xi)+\pi\cot\pi\xi
\end{equation}
in which case we find the condition
\begin{equation}
0=-{1\over a_C}+{2\over a_B}(H(\xi)+\pi\cot\pi\xi)
\end{equation}
in complete agreement with Eq. \ref{eq:bb}.

In the real world---coupled channel---case the analysis is 
similar.  We must generalize the effective Lagrangian to include off
diagonal effects, so that $C_0$ becomes a matrix $(C_0)_{ij}$ with
$i,j=1,2$ and the bubble sum must include two forms of intermediate
states---charged and neutral.  
In the absence of Coulomb effects we have then a set of
equations
\begin{equation}
T_{ij}=(C_0)_{ij}+\sum_{\ell}(C_0)_{i\ell}G_0(k_\ell)T_{\ell j}
\end{equation}
whose solution is 
\begin{equation}
T=(1-C_0G_0(k))^{-1}C_0
\end{equation}
Explicitly, defining 
\begin{equation}
D\equiv (1-(C_0)_{11}G_0(k_1))(1-(C_0)_{22}G_0(k_2))
-(C_0)_{12}(C_0)_{21}G_0(k_1)G_0(k_2)
\end{equation}
we find 
\begin{eqnarray}
T_{11}&=&((C_0)_{11}(1-(C_0)_{22}G_0(k_2))+(C_0)_{12}
(C_0)_{21}G_0(k_2))/D\nonumber\\
T_{21}&=&(C_0)_{21}/D\nonumber\\
T_{12}&=&(C_0)_{12}/D\nonumber\\
T_{22}&=&((C_0)_{22}(1-(C_0)_{11}G_0(k_1))+(C_0)_{12}(C_0)_{21}G_0(k_1))/D
\end{eqnarray}
Using the relation between the scattering amplitude and the T-matrix 
\begin{equation}
f_{ij}=-{\sqrt{m_r^{(i)}m_r^{(j)}}\over 2\pi}T_{ij}
\end{equation}
the connection between the parameters $(C_0)_{ij}$ and the
corresponding scattering lengths $a_{ij}$ is easily found via
\begin{eqnarray}
a_{11}&=&{m_r^{(1)}\over 2\pi}[(C_0(\mu))_{11}-\mu{m_r^{(2)}\over 2\pi}\det
C_0(\mu)]/J\nonumber\\
a_{12}&=&{\sqrt{m_r^{(1)}m_r^{(2)}}\over 2\pi}(C_0(\mu))_{12}/J\nonumber\\
a_{21}&=&{\sqrt{m_r^{(1)}m_r^{(2)}}\over 2\pi}(C_0(\mu))_{21}/J\nonumber\\
a_{22}&=&{m_r^{(2)}\over 2\pi}[(C_0(\mu))_{22}-\mu{m_r^{(1)}\over 2\pi}
\det C_0(\mu)]/J\label{eq:nn}
\end{eqnarray}
with
\begin{equation}
J=1-\mu[{m_r^{(1)}\over 2\pi}(C_0(\mu))_{11}+{m_r^{(2)}\over
2\pi}(C_0(\mu))_{22}]+\mu^2{m_r^{(1)}\over 2\pi}{m_r^{(2)}\over 2\pi}\det C_0(\mu)
\end{equation}
Then Eqs. \ref{eq:nn} are seen to be the same as the K-matrix
forms Eq. \ref{eq:ss}.  Inclusion of the Coulomb interactions is as
before, involving the modifications
\begin{eqnarray}
T&\rightarrow& STS,\quad {\rm where}\quad  S=\left(\begin{array}{cc}
1&0\\
0&C(-\eta_+(k))\exp -i\sigma_0\end{array}\right)\nonumber\\
a_{ij}&\rightarrow&a_{ij}^C\quad{\rm and}\quad -ik_2\rightarrow
{2\over a_B}(H(\xi)+\pi\cot\pi\xi),
\end{eqnarray}
and the resulting bound state condition is identical to Eq. \ref{eq:tt}.
Thus the equivalence of the quantum mechanical and effective
interaction methods is explicitly demonstrated.  

\subsection{Effective Range Effects}

It is straightforward to go to higher order by inclusion of effective range
effects.  In the single channel quantum mechanical formulation this is accomplished
by the modification
\begin{equation}
a_C\rightarrow a_C(1+{1\over 2}a_Cr_Ek^2+\ldots)
\end{equation}
Then Eq. \ref{eq:bb} becomes
\begin{equation}
-{1\over a_C}-{r_E\over 2\xi^2a_B^2}+{2\over
a_B}(H(\xi)+\pi\cot\pi\xi)\simeq0\label{eq:mm}
\end{equation}
whose solution is
\begin{equation}
\Delta E_n=-E_n{4a_C\over na_B}\left[1+{a_C\over a_B}(2H(n)-{3\over
n})+{\cal O}\left(({a_C\over a_B})^2,{a_C r_E\over a_B^2}\right)\right]
\end{equation}
so that effects are negligible.  
Similarly in the two channel case we alter Eq. \ref{eq:rr} via
\begin{equation}
a_{ij}\rightarrow a_{ij}(1+{1\over 4}a_{ij}(r_E)_{ij}(k_i^2+k_j^2)
+\ldots)\label{eq:vvv}
\end{equation}
and the energy shift and width are modified appropriately
\begin{eqnarray}
{\rm Re}\Delta E_n&=&-E_n{4a_{22}^C\over na_B}
\left[1+{a_{22}^C\over a_B}(2H(n)-{3\over n})+{\cal O}\left(({a_{22}^C\over
a_B})^2,{a_{22}^C(r_E)_{22}\over sa_B^2}\right)\right]\nonumber\\
\Gamma_n=2{\rm Im}\Delta E_n&=&-E_n{8\over na_B}k_1|a_{12}^C|^2(1+{1\over
2}a_{12}^C(r_E)_{12}(k_1^2-\kappa_n^2)-(a_{11}^C)^2k_1^2+\ldots)\nonumber\\
\quad
\end{eqnarray}
Again for the real component the effective range effect is tiny and
can generally be neglected.  However, in the case of the decay width
there are two types of corrections which should be included---one due
to the effective range correction and a second due to multiple
rescattering effects in the intermediate state.

Using effective field theory the change is also
directly obtained.  In the single channel case, one 
begins by modifying the effective Lagrangian via
\begin{equation}
{\cal L}\rightarrow{\cal L}+{1\over 2}C_2(\Psi_A^\dagger\vec{\nabla}\Psi_A\cdot
\Psi_B^\dagger\vec{\nabla}\Psi_B+(\vec\nabla\Psi_A^\dagger)\Psi_A\cdot
(\vec\nabla\Psi_B^\dagger)\Psi_B)
\end{equation}
The multiple scattering series may summed exactly as before and the solution
is found as\cite{korea} 
\begin{equation}
T_{fi}(k)=({C_0^R(\mu)+2k^2C_2^R(\mu))C^2(-\eta_+(k))\exp2i\sigma_0
\over 1-(C_0^R(\mu)+2k^2C_2^R(\mu))G_C(k)}
\end{equation}
where $C_0^R(\mu),C_2^R(\mu)$ are the renormalized quantities
\begin{equation}
C_0^R={C_0+LC_2^2\over (1-C_2J)^2}-2{\alpha m_r\mu\over \pi^2}C_2^R,\qquad 
C_2^R={C_2-{1\over 2}JC_2^2\over (1-C_2J)^2}
\end{equation}
and, in cutoff regularization,
\begin{equation}
J=\int{d^3s\over (2\pi)^3}C^2(-\eta_+(s)),\qquad 
L=\int{d^3s\over (2\pi)^3}s^2C^2(-\eta_+(s))
\end{equation}
Thus the scattering amplitude becomes
\begin{eqnarray}
f_C^-(k)&=&-{m_r\over 2\pi}T_{fi}(k)\nonumber\\
&=&{C^2(-\eta_+(k))e^{-2i\sigma_0}
\over -{2\pi\over
m_r(C_0^R(\mu)+2k^2C_2^R(\mu))}-{2\mu\over \pi}+{2\over
a_B}\left[H(i\eta_+(k))-i\pi\coth\pi\eta_+(k))
-\ln{\mu a_B\over 2\pi} -\xi\right]}\nonumber\\
&=&{C^2(-\eta_+(k))e^{-2i\sigma_0}\over -{1\over a_C}+{1\over 2}r_Ek^2
+{2\over a_B}h(\eta_+(k))-ikC^2(-\eta_+(k))}+\ldots
\end{eqnarray}
Thus we identify
\begin{equation}
-{1\over a_C}=-{2\pi\over m_rC_0^R(\mu)}-{2\mu\over \pi}-{2\over
a_B}(\ln{\mu a_B\over 2\pi}+\zeta),\quad 
r_E={8\pi C_2^R(\mu)\over m_r(C_0^{R}(\mu))^2}
\end{equation}
and the bound state condition becomes Eq. \ref{eq:mm}, as expected.
As pointed out in ref.\cite{rk2}, the expression for effective
range is not affected by Coulombic corrections and consequantly 
in the case of the
NN interaction one expects equality for $r_E^{pp}$ and $r_E^{nn}$, as
found experimentally.

The coupled channel analysis is similar---the momentum
dependent effective coupling $C_2$ becomes a $2\times 2$ matrix and
the associated Lippman-Schwinger equation is algebraically somewhat
more complex.  However, the solution is straightforward and amounts in
the end simply to a modification of the effective scattering length parameters
$a_{ij}$ defined in Eq. \ref{eq:nn} by effective range range effects
as given in Eq. \ref{eq:vvv},
where $(r_E)_{ij}$ are given in terms of the matrix elements
$C_2(\mu)_{ij}$.  Thus the
effective interaction bound state condition is identical to Eq.
\ref{eq:oo}, as expected.

\section{Applications}
While the analysis given above is in some ways old, the applications
are not and include two of the most interesting ongoing measurements
in contemporary physics.  One is the problem of
pionium---$\pi^+\pi^-$---whose existence has been claimed by a
Russian group\cite{russ} and which is now being sought in a major program at
CERN.  In this case then the first channel represents $\pi^0\pi^0$ while
the second is $\pi^+\pi^-$.  The reduced mass is 
\begin{equation}
m_r^{(1)}={m_{\pi^0}\over 2},\quad m_r^{(2)}={m_{\pi^+}\over 2}
\end{equation}
and scattering lengths are given in terms of those with total isospin
0,2 via 
\begin{equation}
a_{11}={1\over 3}(2a_2+a_0),\quad a_{12}=a_{21}={\sqrt{2}\over
3}(a_2-a_0),\quad a_{22}={1\over 3}(a_2+2a_0)
\end{equation}
where 
\begin{equation}
a_0=-{7m_\pi\over 32\pi F_\pi^2},\qquad a_2={m_\pi\over 16\pi F_\pi^2}
\end{equation}
are the usual Weinberg values with $F_\pi\simeq 92.4$ MeV being the
pion decay constant.\cite{wein}  To lowest order 
we have a ground state energy shift 
\begin{equation}
\Delta E_{gs}^{(0)}(\pi^+\pi^-)={4\pi\over 3m_{\pi^+}}(a_2+2a_0)|\Psi(0)|^2
\end{equation}
and a width
\begin{equation}
\Gamma_{gs}^{(0)}(\pi^+\pi^-)={16\pi\over 9m_{\pi^+}}\sqrt{2m_{\pi^0}\Delta
m_\pi}(a_2-a_0)^2|\Psi(0)|^2
\end{equation}
which agree with the usual forms.\cite{conv}  Corrections from higher
order effects are found to be
\begin{eqnarray}
{\Delta E_{gs}^{(1)}\over \Delta E_{gs}^{(0)}}&=&{a_{22}\over
a_B}(2H(1)-3)\nonumber\\
{\Gamma_{gs}^{(1)}\over \Gamma_{gs}^{(0)}}&\simeq&(2m_\pi\Delta
m_\pi)({1\over 2}a_{12}(r_E)_{12}-a_{11}^2)
\end{eqnarray}
Using the lowest order chiral symmetry predictions\cite{wein}
\begin{equation}
{1\over 2}a_{12}(r_E)_{12}={4\over 3m_\pi^2},\qquad a_{11}^2
={1\over 32\pi^2F_\pi^2}
\end{equation}
we find a decay rate correction
\begin{equation}
{\Gamma_{gs}^{(1)}\over \Gamma_{gs}^{(0)}}=8.3\%
\end{equation}
in agreement with the result of Kong and Ravndal.\cite{rk1}
\footnote{The dominant effect here is from the effective range,
while the rescattering provides only a small correction.  The form of
our rescattering term differs from that of Kong and Ravndal, as we
include only rescattering from physical---$\pi^0\pi^0$---intermediate
states while they include also that from (unphysical) charged states.} In
the case of the energy shift we find a negligible change
\begin{equation}
{\Delta E_{gs}^{(1)}\over \Delta E_{gs}^{(0)}}=-8.2\times 10^{-4}
\end{equation}
(Strictly speaking, the charged channel phase shift
here---$a_{22}$---should be replaced by its Coulomb corrected value
\begin{equation}
{1\over a_{22}}\rightarrow {1\over a_{22}^C}={1\over a_{22}}+{2\over
a_B}(  
\ln{a_B\over 2R}+1-2\gamma)
\end{equation}
However, this is only a small correction numerically.)  The future
detection of pionium should allow a relatively clean measurement of
the $\pi\pi$ scattering lengths.

The other hadronic atom of current experimental interest is $\pi^-p$
wherein a PSI collaboration has already announced a measurement of the
energy shift and decay rate and which plans to further 
improve these already precise numbers.\cite{psi}  In this case
we are dealing with a three channel system---$\pi^-p,\pi^0n,\gamma n$.
However, both the quantum mechanical calculation as well as 
the effective interaction analysis generalize and the
bound state condition---Eq. \ref{eq:tt}---becomes (after some algebra) 
\begin{eqnarray}
0&=&1+{2\over a_B}(H(\xi)+\pi\cot\pi\xi)\left\{-a_{22}^C\right.\nonumber\\
&+&\left.{ik_1a_{12}^Ca_{21}^C+ik_3a_{32}^Ca_{23}^C-k_1k_3\left[a_{23}^C
(a_{11}^Ca_{32}^C
-a_{31}^Ca_{12}^C)+a_{21}^C(a_{12}^Ca_{33}^C-a_{13}^Ca_{32}^C)\right]\over
1+ik_1a_{11}^C+ik_3a_{33}^C-k_1k_3(a_{11}^Ca_{33}^C-a_{31}^Ca_{13}^C)}\right\}\nonumber\\
\quad\label{eq:kkk}
\end{eqnarray}
To lowest order then we have the results
\begin{eqnarray}
\Delta E_{gs}^{(0)}&=&-E_1{4a_{22}^C\over a_B}\nonumber\\
\Gamma_{gs}^{(0)}&=&-E_1{8\over a_B}(k_1a_{12}^Ca_{21}^C+k_3a_{13}^Ca_{31}^C)
\end{eqnarray}
In the case of the energy shift, the result is as before, but in the
case of the decay width we now have two contributions, one from the
decay to the $\pi^0n$ channel and one from decay to $\gamma n$.
However, the 
$\pi N$ scattering lengths can then be obtained once the radiative
channel is subtracted off via
\begin{equation}
\Gamma_{\pi^0n}={\Gamma_{tot}\over 1+P}
\end{equation}
where $P=\Gamma_{\gamma n}/\Gamma_{\pi^0n}=1.546\pm 0.009$ is 
the Panofsky ratio.\cite{pan}

On the theoretical side the $\pi N$ scattering lengths can be written 
in terms of those with total isospin 
${1\over 2},{3\over 2}$ via
\begin{equation}
a_{11}={1\over 3}(a_1+2a_3),\quad a_{12}=a_{21}={\sqrt{2}\over
3}(a_3-a_1),\quad a_{22}={1\over 3}(2a_1+a_3)
\end{equation}
and the lowest order chiral Lagrangian yields values for these quantities
\begin{equation}
a_1=-2a_3=-{m_\pi\over 4\pi F_\pi^2}{1\over 1+{m_\pi\over m_N}} 
\end{equation}
and higher order chiral corrections have also been
calculated.\cite{meis}  In the case of the radiative channel, the
scattering length is also predicted by chiral symmetry in terms of the
Kroll-Ruderman term\cite{kr}
\begin{equation}
a_{13}=-{\sqrt{2}eg_A\over 8\pi F_\pi}{1\over 1+{m_\pi\over M_N}}
\end{equation}
where $g_A\simeq 1.25$ is the axial coupling in neutron beta decay.

There exist higher order corrections to
these lowest order predictions from both rescattering effects---{\it
cf.} Eq. \ref{eq:kkk}---and from inclusion of effective range
corrections.  However, unlike pionium the latter are not given by chiral
symmetry and hence are model dependent.  Nevertheless when known
corrections are included present experimental results extracted from
$\pi^-p$ experiments
\begin{equation}
b_1={1\over 3}(a_3-a_1)\simeq 0.096\pm 0.007 m_\pi^{-1},\quad b_0\simeq{1\over
3}(a_1+2a_3)=0.0105\pm 0.007 m_\pi^{-1}
\end{equation}
are in satisfactory agreement with the chiral symmetry
prediction\footnote{The isoscalar prediction is more uncertain, as it
has a somewhat senstive dependence on the low energy constants.}
\begin{equation}
0.096m_\pi^{-1}\leq b_1\leq 0.088m_\pi^{-1}
\end{equation}
although there is at present a small discrepancy with a parallel
measurement of the shift in deuterium.\cite{psi}  However, our purpose here is
not to discuss experimental interpretations, rather only to point out
the connection between conventional and effective interaction methods. 

\section{Conclusions}

We have above analyzed the problem of hadronic atom energy shifts due
to strong interaction effects both via a traditional quantum
mechanical discussion and via a calculation 
in the effective interaction picture wherein
the low energy hadronic interactions are written in terms of an
effective local potential and have demonstrated the complete
equivalence between the two procedures.  Applications have been made
to the systems pionium---$\pi^+\pi^-$---and $\pi^-p$ currently being
studied experimentally at CERN and PSI respectively.

\begin{center}
Acknowledgement
\end{center}

It is a pleasure to acknowledge very helpful communications with 
Finn Ravndal as well
as the support of the Alexander von Humboldt Foundation and the hospitality of 
Forschungszentrum J\"{u}lich.  This work was also supported in 
part by the National Science Foundation.

\end{document}